\documentclass[12pt]{article}
\usepackage{makeidx}
\usepackage{amssymb}
\usepackage{amsfonts}
\usepackage{amsmath}
\usepackage{graphicx}
\usepackage{setspace}
\usepackage{authblk}
\usepackage{units}
\usepackage{appendix}
\newcommand{\lambdabar}{{\mkern0.75mu\mathchar '26\mkern -9.75mu\lambda}}
\usepackage[left=2cm,top=2cm,right=2cm,bottom=2cm]{geometry}

\begin{document}

\title{\textbf{Role of switching-on and -off effects in the vacuum instability}}
\author[1,2]{T. C. Adorno\thanks{tg.adorno@gmail.com, tg.adorno@mail.tsu.ru}}
\author[5]{R. Ferreira\thanks{rafaelufpi@gmail.com}}
\author[1,3]{S. P. Gavrilov\thanks{gavrilovsergeyp@yahoo.com, gavrilovsp@herzen.spb.ru}}
\author[1,4,5]{D. M. Gitman\thanks{gitman@if.usp.br}}
\affil[1]{\textit{Department of Physics, Tomsk State University, Lenin Prospekt 36, 634050, Tomsk, Russia;}}
\affil[2]{\textit{College of Physical Science and Technology, Hebei University, Wusidong Road 180, 071002, Baoding, China;}}
\affil[3]{\textit{Department of General and Experimental Physics, Herzen State Pedagogical University of Russia, Moyka embankment 48, 191186, St. Petersburg, Russia;}}
\affil[4]{\textit{P. N. Lebedev Physical Institute, 53 Leninskiy prospekt,
119991, Moscow, Russia;}}
\affil[5]{\textit{Instituto de F\'{\i}sica, Universidade de S\~{a}o Paulo, Caixa Postal 66318, CEP 05508-090, S\~{a}o Paulo, S.P., Brazil;}}

\maketitle

\onehalfspacing

\begin{abstract}
We find exact differential mean numbers of fermions and bosons created from the vacuum due to a composite electric field of special configuration. This configuration imitates a finite switching-on and -off regime and consists of fields that switch-on exponentially from the infinitely remote past, remains constant during a certain interval $T$ and switch-off exponentially to the infinitely remote future. We show that calculations in the slowly varying field approximation are completely predictable in the framework of a locally constant field approximation. Beyond the slowly varying field approximation, we study effects of fast switching-on and -off in a number of cases when the size of the dimensionless parameter  $\sqrt{eE}T$ is either close or exceeds the threshold value that determines the transition from a regime sensitive to on-off parameters to the slowly varying regime for which
these effects are secondary.


Particle creation, Schwinger effect, time-dependent external field, Dirac and Klein-Gordon equations.
\end{abstract}

\section{Introduction\label{Sec1}}

The consistent consideration of quantum processes in the vacuum violating
backgrounds has to be done in the framework of nonperturbative calculations
for quantum field theory, in particular, QED. Different analytical and
numerical methods were applied to study the effect of electron-positron pair
creation from vacuum; see recent reviews \cite{Ruffini,Gelis}. Among these
methods, there are ones based on the existence of exact solutions of the
Dirac (or Klein-Gordon) equation in the corresponding background fields;
e.g., see Refs. \cite{FGS,GavGit16}. They give us exactly solvable models
for QED that are useful to consider the characteristic features of theory
and could be used to check approximations and numeric calculations.
Recently, we present the review of particle creation effects in
time-dependent uniform external electric fields that contains three most
important exactly solvable cases: Sauter-like electric field, $T$-constant
electric field, and exponentially growing and decaying electric fields \cite%
{AdoGavGit17}. These electric fields switched on and off at the initial and
the final time instants, respectively. We refer to such kind of external
fields as the $t$-electric potential steps.

Choosing parameters for the exponentially varying electric fields, one can
consider both fields in the slowly varying regime and fields that exist only
for a short time in a vicinity of the switching on and off time. The case of
the $T$-constant electric field is distinct. In this case the electric field
is constant within the time interval $T$ and is zero outside of it, that is,
it is switched on and off ``abruptly'' at
definite instants. The model with the $T$-constant electric field is
important to study the particle creation effects; see Ref. \cite{AdoGavGit17}
for the review. Then the details of switching on and off for the $T$%
-constant electric field is of interest. To estimate the role of the
switching on and off effects for the pair creation due to the $T$-constant
electric field we consider a composite electric field that grows
exponentially in the first interval $t\in \mathrm{I}=\left( -\infty
,t_{1}\right) $, remains constant in the second interval $t\in \mathrm{II}=%
\left[ t_{1},t_{2}\right] $,$\ $and decreases exponentially in the last
interval $t\in \mathrm{III}=\left( t_{2},+\infty \right) $. We essentially
use notation and final formulas from Ref. \cite{AdoGavGit17}.

The article is organized as follows: In Sec. \ref{Sec2}, we introduce the
composite field and summarize details concerning the exact solutions of the
Dirac equation with such a field. We find exact formulas for the
differential mean number of particles created from the vacuum, the total
number of particles created from the vacuum and the vacuum-to-vacuum
transition probability. In Sec. \ref{Sec3} we consider the general
properties of the differential mean numbers of pairs created. We visualize
how these mean numbers are distributed over the quantum numbers, especially
in cases where asymptotic approximations involved are not applicable. In
Sec. \ref{Sec4} we compute differential and total quantities in some special
field configurations of interest. We show that the results for slowly
varying fields are completely predictable using recently developed version
of a locally constant field approximation. We study configurations that
simulate finite switching-on and -off processes within and beyond the slowly
varying regime. Final comments are placed in Sec. \ref{conclusions}.

\section{IN and OUT solutions in a composite electric field\label{Sec2}}

In this section we summarize general aspects on exact solutions of the Dirac
equation with the field under consideration and briefly discuss the
calculation of differential and total numbers of pairs creation.

The composite electric field a $d=D+1$ dimensional Minkowski space-time is
homogeneous, positively oriented along a single direction $\mathbf{E}\left(
t\right) =\left( E^{i}\left( t\right) =\delta _{1}^{i}E\left( t\right) \,,\
\ i=1,...,D\right) $ and described by a vector along the same direction, $%
A^{\mu }=\left( A^{0}=0,\mathbf{A}\left( t\right) \right) $, $\mathbf{A}%
\left( t\right) =\left( A^{i}\left( t\right) =\delta _{1}^{i}A_{x}\left(
t\right) \right) $, whose explicit forms are%
\begin{eqnarray}
&&E\left( t\right) =E\left\{
\begin{array}{ll}
e^{k_{1}\left( t-t_{1}\right) }\,, & t\in \mathrm{I}\,, \\
1\,, & t\in \mathrm{II\,}, \\
e^{-k_{2}\left( t-t_{2}\right) }\,, & t\in \mathrm{III\,},%
\end{array}%
\right. \ \ \left( E,k_{1},k_{2}\right) >0\,,  \label{s2.0} \\
&&A_{x}\left( t\right) =E\left\{
\begin{array}{ll}
k_{1}^{-1}\left( -e^{k_{1}\left( t-t_{1}\right) }+1-k_{1}t_{1}\right) \,, &
t\in \mathrm{I}\,, \\
-t\,, & t\in \mathrm{II\,}, \\
k_{2}^{-1}\left( e^{-k_{2}\left( t-t_{2}\right) }-1-k_{2}t_{2}\right) \,, &
t\in \mathrm{III}\,,%
\end{array}%
\right.  \label{s2.1}
\end{eqnarray}%
where $t_{1}<0$ and $t_{2}>0$ are fixed time instants. Throughout the text,
we refer to \textrm{I} as the switching-on interval, \textrm{III} as the
switching-off interval and \textrm{II} as the constant field interval.
This field configuration encompasses the $T$\ -constant field \cite{GavGit96}, characterized by the
absence of exponential parts, and the peak field \cite{AdoGavGit16}.

The Dirac equation\footnote{%
The subscript \textquotedblleft $\perp $\textquotedblright\ denotes spacial
components perpendicular to the electric field (e. g. $\mathbf{x}_{\perp
}=\left\{ x^{2},...,x^{D}\right\} $) and $\psi (x)$ is a $2^{[d/2]}$%
-component spinor ($[d/2]$ stands for the integer part of the ratio $d/2$).
As usual, $m$ denotes the electron mass, $\gamma ^{\mu }$ are $\gamma $%
-matrices in $d$ dimensions and $U\left( t\right) $ denotes the potential
energy of a particle with algebraic charge $q$. We select the electron as
the main particle, $q=-e$ with $e$ representing the absolute value of the
electron charge. Hereafter we use the relativistic system of units ($\hslash
=c=1$), except when indicated otherwise.}%
\begin{eqnarray}
&&i\partial _{t}\psi \left( x\right) =H\left( t\right) \psi \left( x\right)
\,,\ \ H\left( t\right) =\gamma ^{0}\left( \boldsymbol{\gamma }\mathbf{P}%
+m\right) \,,  \notag \\
&&\,P_{x}=-i\partial _{x}-U\left( t\right) ,\ \ \mathbf{P}_{\bot }=-i%
\boldsymbol{\nabla }_{\perp },\ \ U\left( t\right) =qA_{x}\left( t\right) \,,
\label{s3}
\end{eqnarray}%
can be solved exactly in each one of the intervals above. Once the
corresponding exact solutions are known (see, e.g., the review \cite{AdoGavGit17}), we only
present few details to obtain such a solutions. Firstly, we represent the
Dirac spinors $\psi _{n}\left( x\right) $ in terms of new time-dependent
spinors $\phi _{n}(t)$ as%
\begin{eqnarray}
&&\psi _{n}\left( x\right) =\exp \left( i\mathbf{pr}\right) \psi _{n}\left(
t\right) \,,\ \ n=(\mathbf{p},\sigma )\,,  \notag \\
&&\psi _{n}\left( t\right) =\left\{ \gamma ^{0}i\partial _{t}-\left[
p_{x}-U\left( t\right) \right] -\boldsymbol{\gamma }\mathbf{p}+m\right\}
\phi _{n}(t)\,,  \label{2.10}
\end{eqnarray}%
and separate the spinning degrees of freedom by the substitution $\phi
_{n}(t)=\varphi _{n}\left( t\right) v_{\chi ,\sigma }$, in which $v_{\chi
,\sigma }$ and $\varphi _{n}\left( t\right) $ denotes a set of constant
orthonormalized spinors and scalar functions, respectively. The constant
spinors satisfy
\begin{equation}
\gamma ^{0}\gamma ^{1}v_{\chi ,\sigma }=\chi v_{\chi ,\sigma }\,,\ \ v_{\chi
,\sigma }^{\dag }v_{\chi ^{\prime },\sigma ^{\prime }}=\delta _{\chi ,\chi
^{\prime }}\delta _{\sigma ,\sigma ^{\prime }\,},  \label{e2a}
\end{equation}%
where $\chi =\pm 1$ are eigenvalues of $\gamma ^{0}\gamma ^{1}$ and $\sigma
=(\sigma _{1},\sigma _{2},\dots ,\sigma _{\lbrack d/2]-1})$ represent a set
of additional eigenvalues, corresponding to spin operators compatible with $%
\gamma ^{0}\gamma ^{1}$. The constant spinors are subjected to additional
conditions depending on the space-time dimensions, whose details can be
found in Ref. \cite{AdoGavGit17}. After these substitutions, the Dirac
spinor can be obtained through the solutions of the second-order ordinary
differential equation\footnote{%
For scalar particles, the exact solutions for the Klein-Gordon equation $%
\phi _{n}\left( x\right) $ are connected with the scalar functions as $\phi
_{n}\left( x\right) =\exp \left( i\mathbf{pr}\right) \varphi _{n}\left(
t\right) $. Since spinning degrees-of-freedom are absent in this case, $n=%
\mathbf{p}$ and $\chi =0$ in Eq. (\ref{s2}) as well as in all subsequent
formulas.}%
\begin{equation}
\left\{ \frac{d^{2}}{dt^{2}}+\left[ p_{x}-U\left( t\right) \right] ^{2}+\pi
_{\perp }^{2}-i\chi \dot{U}\left( t\right) \right\} \varphi _{n}\left(
t\right) =0\,,\ \ \pi _{\perp }=\sqrt{\mathbf{p}_{\perp }^{2}+m^{2}}\,.
\label{s2}
\end{equation}%
In the switching-on \textrm{I} and -off \textrm{III} intervals, the
solutions are expressed in terms of Confluent Hypergeometric Functions
(CHFs.),%
\begin{align}
& \varphi _{n}^{j}\left( t\right) =b_{2}^{j}y_{1}^{j}\left( \eta _{j}\right)
+b_{1}^{j}y_{2}^{j}\left( \eta _{j}\right) \,,  \notag \\
& y_{1}^{j}\left( \eta _{j}\right) =e^{-\eta _{j}/2}\eta _{j}^{\nu _{j}}\Phi
\left( a_{j},c_{j};\eta _{j}\right) \,,  \notag \\
& y_{2}^{j}\left( \eta _{j}\right) =e^{\eta _{j}/2}\eta _{j}^{-\nu _{j}}\Phi
\left( 1-a_{j},2-c_{j};-\eta _{j}\right) \,,  \label{i.3.3}
\end{align}%
while at the constant interval \textrm{II}, the solutions are expressed in
terms of Weber Parabolic Cylinder Functions (WPCFs.),%
\begin{eqnarray}
\varphi _{n}\left( z\right) &=&b^{+}u_{+}\left( z\right) +b^{-}u_{-}\left(
z\right) \,,  \notag \\
u_{+}\left( z\right) &=&D_{\beta +\left( \chi -1\right) /2}\left( z\right)
\,,\ \ u_{-}\left( z\right) =D_{-\beta -\left( \chi +1\right) /2}\left(
iz\right) \,.  \label{ii.5}
\end{eqnarray}%
At these equations, $a_{j}$, $c_{j}$, $\nu _{j}$ and $\beta $ are parameters%
\begin{eqnarray}
&&a_{1}=\frac{1}{2}\left( 1+\chi \right) +i\Xi _{1}^{-}\,,\ \ a_{2}=\frac{1}{%
2}\left( 1+\chi \right) +i\Xi _{2}^{+}\,,  \notag \\
&&\Xi _{j}^{\pm }=\frac{\omega _{j}\pm \Pi _{j}}{k_{j}}\,,\ \ c_{j}=1+2\nu
_{j}\,,\ \ \nu _{j}=\frac{i\omega _{j}}{k_{j}}\,,\ \ \beta =\frac{i\lambda }{%
2}\,,  \notag \\
&&\omega _{j}=\sqrt{\Pi _{j}^{2}+\pi _{\perp }^{2}}\,,\ \ \Pi _{j}=p_{x}-%
\frac{eE}{k_{j}}\left[ \left( -1\right) ^{j}+k_{j}t_{j}\right] \,,\ \
\lambda =\frac{\pi _{\perp }^{2}}{eE}\,,  \label{i.3}
\end{eqnarray}%
$z$ and $\eta _{j}$ are time-dependent functions%
\begin{eqnarray}
&&\eta _{1}\left( t\right) =ih_{1}e^{k_{1}\left( t-t_{1}\right) }\,,\ \ \eta
_{2}\left( t\right) =ih_{2}e^{-k_{2}\left( t-t_{2}\right) }\,,\ \ h_{j}=%
\frac{2eE}{k_{j}^{2}}\,,  \label{i.0} \\
&&z\left( t\right) =\left( 1-i\right) \xi \left( t\right) \,,\ \ \xi \left(
t\right) =\frac{eEt-p_{x}}{\sqrt{eE}}\,,  \label{ii.3}
\end{eqnarray}%
and $b_{1,2}^{j}$, $b^{\pm }$ are constants, fixed by initial conditions. In
addition, the index $j$ in Eqs. (\ref{i.3.3}), (\ref{i.3}) and (\ref{i.0})
distinguish quantities associated to the switching-on $\left( j=1\right) $
from the switching-off $\left( j=2\right) $ intervals.

In virtue of asymptotic properties of the CHFs. at $t\rightarrow \pm \infty $%
, the solutions given by Eq. (\ref{i.3.3}) can be classified as
particle/antiparticle states%
\begin{eqnarray}
\ _{+}\varphi _{n}\left( t\right) &=&\ _{+}\mathcal{N}\exp \left( i\pi \nu
_{1}/2\right) y_{2}^{1}\left( \eta _{1}\right) \,,\,\ _{-}\varphi _{n}\left(
t\right) =\ _{-}\mathcal{N}\exp \left( -i\pi \nu _{1}/2\right)
y_{1}^{1}\left( \eta _{1}\right) \,,\ \ t\in \mathrm{I}\,,  \notag \\
\ ^{+}\varphi _{n}\left( t\right) &=&\ ^{+}\mathcal{N}\exp \left( -i\pi \nu
_{2}/2\right) y_{1}^{2}\left( \eta _{2}\right) \,,\,\ ^{-}\varphi _{n}\left(
t\right) =\ ^{-}\mathcal{N}\exp \left( i\pi \nu _{2}/2\right)
y_{2}^{2}\left( \eta _{2}\right) \,,\ \ t\in \mathrm{III}\,,  \label{i.4.1}
\end{eqnarray}%
since, at the infinitely remote past $t\rightarrow -\infty $ and future $%
t\rightarrow +\infty $, the set above behaves as plane-waves,%
\begin{equation}
\ _{\zeta }\varphi _{n}\left( t\right) =\ _{\zeta }\mathcal{N}e^{-i\zeta
\omega _{1}t}\,,\ \ t\rightarrow -\infty \,,\ \ ^{\zeta }\varphi _{n}\left(
t\right) =\ ^{\zeta }\mathcal{N}e^{-i\zeta \omega _{2}t}\,,\ \ t\rightarrow
+\infty \,,  \label{i.4.0}
\end{equation}%
where $\omega _{1}$ denotes the energy of initial particles at $t\rightarrow
-\infty $, $\omega _{2}$ denotes the energy of final particles at $%
t\rightarrow +\infty $ and $\zeta $ labels electron $\left( \zeta =+\right) $
and positron $\left( \zeta =-\right) $ states. With the help of such
solutions, one may construct IN $\left\{ \ _{\zeta }\psi \left( x\right)
\right\} $ and OUT $\left\{ \ ^{\zeta }\psi \left( x\right) \right\} $ sets
of Dirac spinors. The normalization constants$\;_{\zeta }\mathcal{N}=\
_{\zeta }CV_{\left( d-1\right) }^{-1/2}$ and $\;^{\zeta }\mathcal{N}=\
^{\zeta }CV_{\left( d-1\right) }^{-1/2}$ are calculated with respect to the
usual inner product for Fermions and Bosons, where $\ _{\zeta }C$ and $\
^{\zeta }C$ given by%
\begin{equation}
\ _{\zeta }C=\left\{
\begin{array}{ll}
\left( 2\omega _{1}q_{1}^{\zeta }\right) ^{-1/2}\,, & \mathrm{Fermi\,,} \\
\left( 2\omega _{1}\right) ^{-1/2}\,, & \mathrm{Bose\,,}%
\end{array}%
\right. \,,\ ^{\zeta }C=\left\{
\begin{array}{ll}
\left( 2\omega _{2}q_{2}^{\zeta }\right) ^{-1/2}\,, & \mathrm{Fermi\,,} \\
\left( 2\omega _{2}\right) ^{-1/2}\,, & \mathrm{Bose\,,}%
\end{array}%
\right. \,,\ q_{j}^{\zeta }=\omega _{j}-\chi \zeta \Pi _{j}\,.  \label{i.4.2}
\end{equation}%
For further details, e.g., see Ref. \cite{AdoGavGit17}.

With the exact solutions discussed above, one can write complete sets of
solutions for the whole time interval $t\in \left( -\infty ,+\infty \right) $%
. Using the classification (\ref{i.4.1}) and the solutions given by Eq. (\ref%
{ii.5}), Dirac spinors (\ref{2.10}) (or Klein-Gordon solutions) for all time
duration can be calculated from the following set of solutions,%
\begin{eqnarray}
\ ^{+}\varphi _{n}\left( t\right) &=&\left\{
\begin{array}{ll}
\kappa g\left( _{-}|^{+}\right) \ _{-}\varphi _{n}\left( t\right) +g\left(
_{+}|^{+}\right) \ _{+}\varphi _{n}\left( t\right) \,, & t\in \mathrm{I}\,
\\
b_{1}^{+}u_{+}\left( t\right) +b_{1}^{-}u_{-}\left( t\right) \,, & t\in
\mathrm{II\,} \\
\ ^{+}\mathcal{N}\exp \left( -i\pi \nu _{2}/2\right) y_{1}^{2}\left( \eta
_{2}\right) \,, & t\in \mathrm{III\,}%
\end{array}%
\right. ;  \label{v1} \\
\ _{-}\varphi _{n}\left( t\right) &=&\left\{
\begin{array}{ll}
\;_{-}\mathcal{N}\exp \left( -i\pi \nu _{1}/2\right) y_{1}^{1}\left( \eta
_{1}\right) \,, & t\in \mathrm{I\,} \\
b_{2}^{+}u_{+}\left( t\right) +b_{2}^{-}u_{-}\left( t\right) \,, & t\in
\mathrm{II\,} \\
g\left( ^{+}|_{-}\right) \ ^{+}\varphi _{n}\left( t\right) +\kappa g\left(
^{-}|_{-}\right) \ ^{-}\varphi _{n}\left( t\right) \,, & t\in \mathrm{III\,}%
\end{array}%
\right. ,  \label{v4}
\end{eqnarray}%
where $b_{1,2}^{\pm }$, $g\left( _{\pm }|^{+}\right) $, and $g\left( ^{\pm
}|_{-}\right) $ are some coefficients, $g\left( ^{\zeta ^{\prime }}|_{\zeta
}\right) =g\left( _{\zeta ^{\prime }}|^{\zeta }\right) ^{\ast }$. Here $%
\kappa $ is an auxiliary constant that allow us to present solutions for the
Klein-Gordon $\left( \kappa =-1\right) $ or Dirac $\left( \kappa =+1\right) $
equations. For the solutions of the Dirac equation, the $g$-coefficients
satisfy unitarity relations%
\begin{equation}
\sum_{\varkappa }g\left( ^{\zeta }|_{\varkappa }\right) g\left( _{\varkappa
}|^{\zeta ^{\prime }}\right) =\sum_{\varkappa }g\left( _{\zeta }|^{\varkappa
}\right) g\left( ^{\varkappa }|_{\zeta ^{\prime }}\right) =\delta _{\zeta
,\zeta ^{\prime }}\,  \label{v4.2}
\end{equation}%
while for the solutions of the Klein-Gordon equation, the $g$-coefficients
satisfy unitarity relations

\begin{equation}
\sum_{\varkappa }\varkappa g\left( ^{\zeta }|_{\varkappa }\right) g\left(
_{\varkappa }|^{\zeta ^{\prime }}\right) =\sum_{\varkappa }\varkappa g\left(
_{\zeta }|^{\varkappa }\right) g\left( ^{\varkappa }|_{\zeta ^{\prime
}}\right) =\zeta \delta _{\zeta ,\zeta ^{\prime }}\,,  \label{v4.4}
\end{equation}

To obtain the $g$-coefficients, we conveniently consider continuity
conditions at instants $t_{1}$, $t_{2}$%
\begin{equation*}
\ _{-}^{+}\varphi _{n}\left( t_{1,2}-0\right) =\ _{-}^{+}\varphi _{n}\left(
t_{1,2}+0\right) \,,\ \ \partial _{t}\ _{-}^{+}\varphi _{n}\left(
t_{1,2}-0\right) =\partial _{t}\ _{-}^{+}\varphi _{n}\left( t_{1,2}+0\right)
\,,
\end{equation*}%
substitute appropriate normalization constants for each case, given by Eqs. (%
\ref{i.4.2}), and use Wronskian determinants for CHFs. and WPCFs. After
these manipulations, one can readily verify that $g\left( _{-}|^{+}\right) $
and $g\left( ^{+}|_{-}\right) $ for the Dirac case reads%
\begin{eqnarray}
g\left( _{-}|^{+}\right) &=&\sqrt{\frac{q_{1}^{-}}{8eE\omega
_{1}q_{2}^{+}\omega _{2}}}\exp \left[ \frac{i\pi }{2}\left( \nu _{1}-\nu
_{2}+\beta +\frac{\chi }{2}\right) \right] \left[ f_{1}^{-}\left(
t_{2}\right) f_{2}^{+}\left( t_{1}\right) -f_{1}^{+}\left( t_{2}\right)
f_{2}^{-}\left( t_{1}\right) \right] \,,  \notag \\
g\left( ^{+}|_{-}\right) &=&\sqrt{\frac{q_{2}^{+}}{8eE\omega
_{2}q_{1}^{-}\omega _{1}}}\exp \left[ \frac{i\pi }{2}\left( \nu _{2}-\nu
_{1}+\beta +\frac{\chi }{2}\right) \right] \left[ f_{1}^{+}\left(
t_{1}\right) f_{2}^{-}\left( t_{2}\right) -f_{1}^{-}\left( t_{1}\right)
f_{2}^{+}\left( t_{2}\right) \right] \,,  \notag \\
f_{k}^{\pm }\left( t_{j}\right) &=&\left. \left[ (-1)^{j}k_{j}\eta _{j}\frac{%
dy_{k}^{j}\left( \eta _{j}\right) }{d\eta _{j}}+y_{k}^{j}\left( \eta
_{j}\right) \partial _{t}\right] u_{\pm }\left( z\right) \right\vert
_{t=t_{j}}\,,  \label{r4}
\end{eqnarray}%
while for the Klein-Gordon case have the form,%
\begin{eqnarray}
g\left( _{-}|^{+}\right) &=&-\frac{1}{\sqrt{8eE\omega _{1}\omega _{2}}}\exp %
\left[ \frac{i\pi }{2}\left( \nu _{1}-\nu _{2}+\beta \right) \right] \left. %
\left[ f_{1}^{-}\left( t_{2}\right) f_{2}^{+}\left( t_{1}\right)
-f_{1}^{+}\left( t_{2}\right) f_{2}^{-}\left( t_{1}\right) \right]
\right\vert _{\chi =0}\,,  \notag \\
g\left( ^{+}|_{-}\right) &=&\frac{1}{\sqrt{8eE\omega _{1}\omega _{2}}}\exp %
\left[ \frac{i\pi }{2}\left( \nu _{2}-\nu _{1}+\beta \right) \right] \left. %
\left[ f_{1}^{+}\left( t_{1}\right) f_{2}^{-}\left( t_{2}\right)
-f_{1}^{-}\left( t_{1}\right) f_{2}^{+}\left( t_{2}\right) \right]
\right\vert _{\chi =0}\,.  \label{r7}
\end{eqnarray}

Taking into account that the $g$'s coefficients establish the Bogoliubov
transformations, one may compute fundamental quantities concerning vacuum
instability for Fermions (the Dirac case) and Bosons (the Klein-Gordon
case), for example, the differential mean number of pairs created from the
vacuum $N_{n}^{\mathrm{cr}}$, the total number $N$ and the vacuum-to-vacuum
transition probability $P_{v}$ as%
\begin{eqnarray}
&&N_{n}^{\mathrm{cr}}=\left\vert g\left( _{-}|^{+}\right) \right\vert
^{2}\,,\ \ N^{\mathrm{cr}}=\sum_{n}N_{n}^{\mathrm{cr}}\,,  \notag \\
&&P_{v}=\exp \left[ \kappa \sum_{n}\ln \left( 1-\kappa N_{n}^{\mathrm{cr}%
}\right) \right] \,.  \label{NP}
\end{eqnarray}

\section{General properties of the differential mean numbers of pairs
created \label{Sec3}}

The $g$-coefficients (\ref{r4}) and (\ref{r7}) enjoy certain properties
under time/momentum reversal that result in symmetries for differential
quantities. More precisely, the simultaneous change%
\begin{equation}
k_{1}\leftrightarrows k_{2}\,,\ \ t_{1}\leftrightarrows -t_{2}\,,\ \
p_{x}\leftrightarrows -p_{x}\,,  \label{sym1}
\end{equation}%
yields to a number of identities, for instance, $\Pi _{1}\leftrightarrows
-\Pi _{2}$, $\omega _{1}\leftrightarrows \omega _{2}$, $a_{1}%
\leftrightarrows a_{2}$, $c_{1}\leftrightarrows c_{2}$ so that $g\left(
_{-}|^{+}\right) $ and $g\left( ^{+}|_{-}\right) $ are related by%
\begin{equation}
g\left( _{-}|^{+}\right) \leftrightarrows \kappa g\left( ^{+}|_{-}\right) \,,
\label{sym2}
\end{equation}%
implying, in particular, that $N_{n}^{\mathrm{cr}}$ (and therefore total
quantities) are even with respect to the exchanges (\ref{sym1}). Moreover (%
\ref{r4}) and (\ref{r7}) are even with respect to $\mathbf{p}_{\perp }$, so
that all quantum quantities in Eq. (\ref{NP}) are symmetric with respect to
the momenta $\mathbf{p}$ (for Fermions, these quantities does not depend on
spin polarizations as well). Such properties are helpful in computing
asymptotic estimates in several regimes, some of them discussed in
subsequent section.

\begin{figure}[th]
\begin{center}
\includegraphics[scale=0.48]{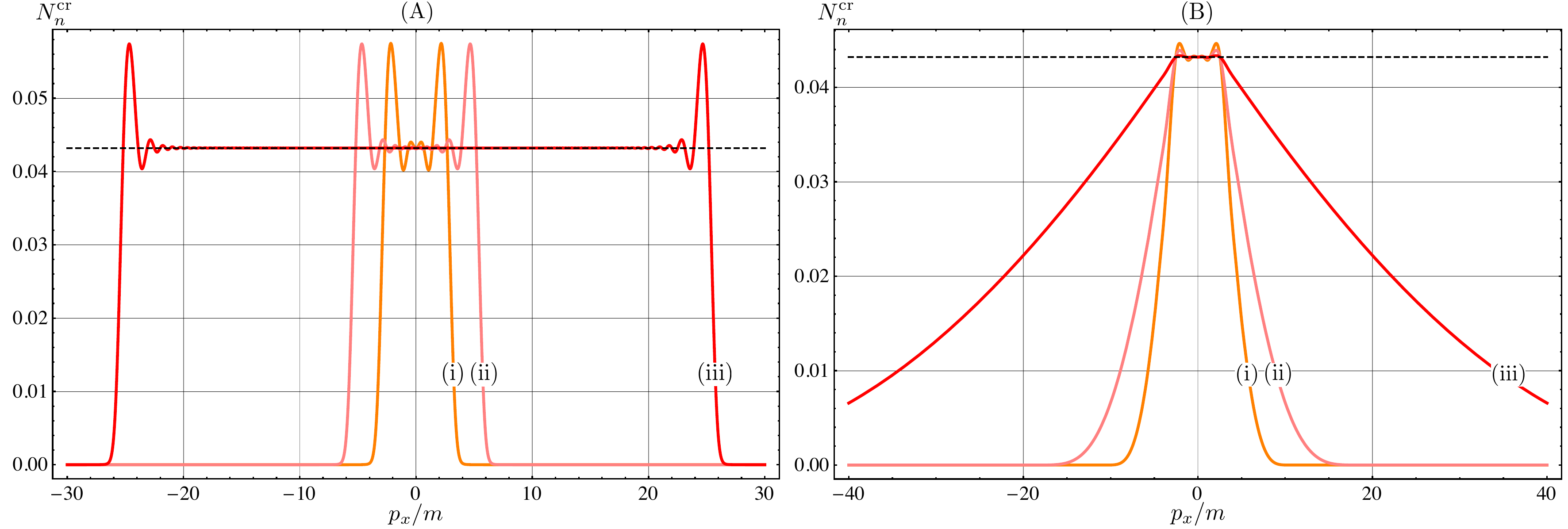}
\end{center}
\caption{(color online) Differential mean number of Fermions created from
the vacuum $N_{n}^{\mathrm{cr}}$ (solid lines) by a symmetrical composite
field, with $k_{1}=k_{2}=k$ and amplitude $E=E_{\mathrm{c}}=m^{2}/e=1$
fixed. Graph (A) shows distributions with $k/m=1$ fixed,
while Graph (B) shows $mT$ $\ $ is fixed, $mT=5$. In (A), the solid lines labeled with $(\mathrm{i})$, $(\mathrm{ii})$ and $(\mathrm{iii})$ refers to $mT=5$, $mT=10$ and $mT=50$, respectively. In (B), $(\mathrm{i})$, $(\mathrm{ii})$ and $(\mathrm{iii})$ refers to $k/m=0.1$, $k/m=0.05$ and $k/m=0.01$, respectively. The horizontal dashed line corresponds to the uniform distribution $e^{-\protect
\pi \protect\lambda }$ which, in this system of units and $\mathbf{p}%
_{\perp }=0$, is $e^{-\protect\pi }$.}
\label{Fig1a}
\end{figure}

\begin{figure}[th]
\begin{center}
\includegraphics[scale=0.48]{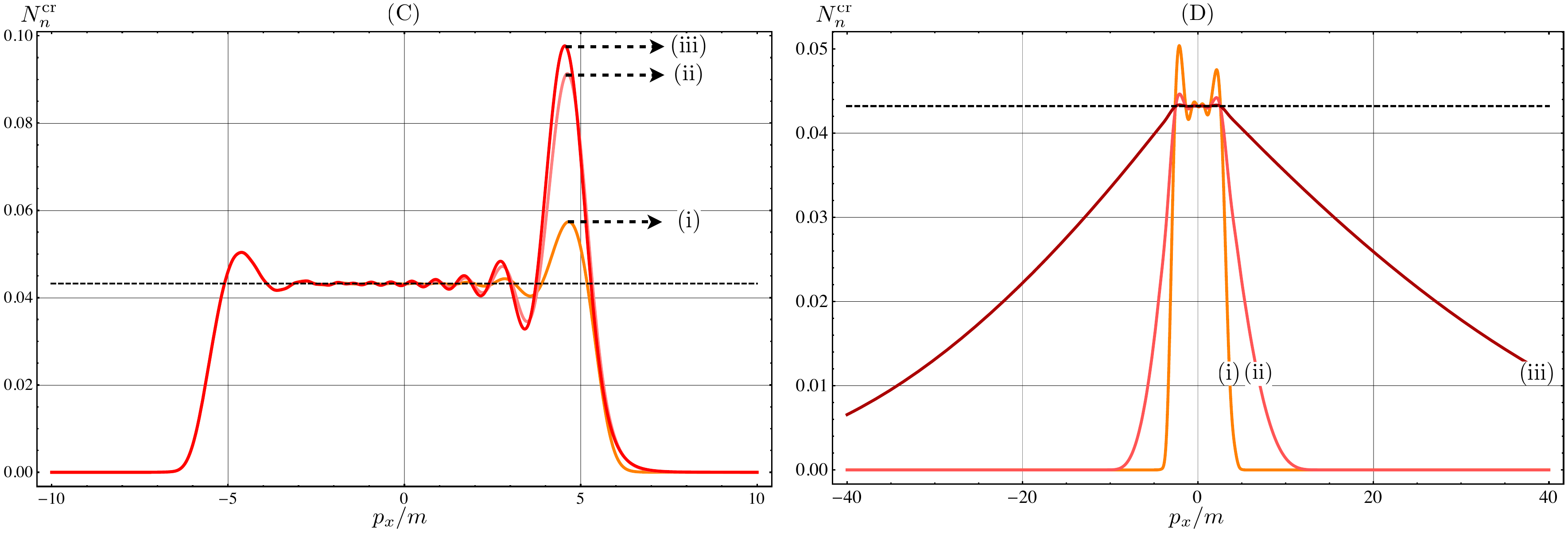}
\end{center}
\caption{(color online) Differential mean number of Fermions created from
the vacuum $N_{n}^{\mathrm{cr}}$ (solid lines) by asymmetrical composite
fields, with $k_{1}\neq k_{2}$ and amplitude $E=E_{\mathrm{c}}=m^{2}/e=1$
fixed. In both graphs, $mT$ is fixed, where in Graph
(C) shows $mT=10$ and $k_{1}/m=0.5$ while in Graph (D) shows $mT=5$.
In (C), the solid lines labeled with $(\mathrm{i})$, $(\mathrm{ii})$ and $(\mathrm{iii})$ refers to $k_2/m=1$, $k_2/m=5$ and $k_2/m=10$, respectively. In (D), $(\mathrm{i})$ denotes $k_1/m=0.5,\,k_2/m=0.3$, $(\mathrm{ii})$ denotes $k_1/m=0.1,\,k_2/m=0.07$ and $(\mathrm{iii})$ denotes $k_1/m=0.01,\,k_2/m=0.008$.
The horizontal dashed line corresponds to the uniform distribution $e^{-%
\protect\pi \protect\lambda }$ which, in this system of units and $%
\mathbf{p}_{\perp }=0$, is $e^{-\protect\pi }$.}
\label{Fig1b}
\end{figure}

\begin{figure}[th]
\begin{center}
\includegraphics[scale=0.48]{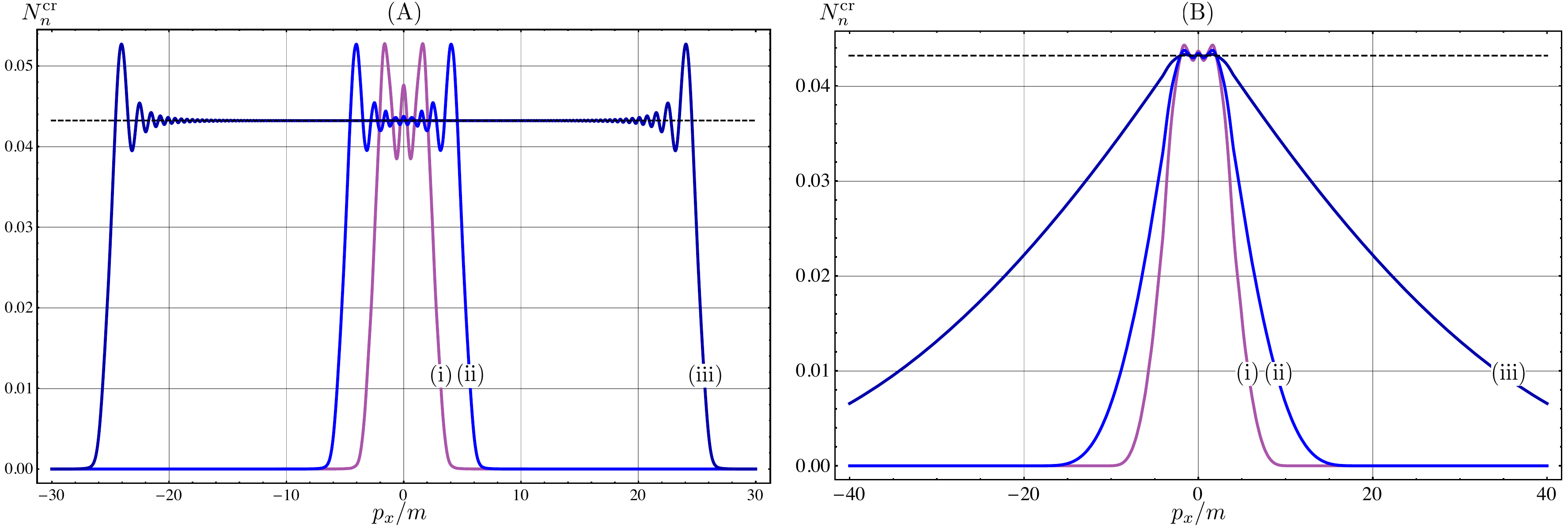}
\end{center}
\caption{(color online) Differential mean number of Bosons created from the
vacuum $N_{n}^{\mathrm{cr}}$ (solid lines) by a symmetrical composite field,
with $k_{1}=k_{2}=k$ and amplitude $E=E_{\mathrm{c}}=m^{2}/e=1$ fixed. 
Graph (A) shows distributions with $k/m=1$ fixed,
while Graph (B) shows $mT$ is fixed, $mT=5$. In (A), the solid lines labeled with $(\mathrm{i})$, $(\mathrm{ii})$ and $(\mathrm{iii})$ refers to $mT=5$, $mT=10$ and $mT=50$, respectively. In (B), $(\mathrm{i})$, $(\mathrm{ii})$ and $(\mathrm{iii})$ refers to $k/m=0.1$, $k/m=0.05$ and $k/m=0.01$, respectively. The horizontal dashed line corresponds to the uniform distribution $e^{-\protect\pi \protect%
\lambda }$ which, in this system of units and $\mathbf{p}_{\perp }=0$%
, is $e^{-\protect\pi }$.}
\label{Fig2a}
\end{figure}

\begin{figure}[th]
\begin{center}
\includegraphics[scale=0.48]{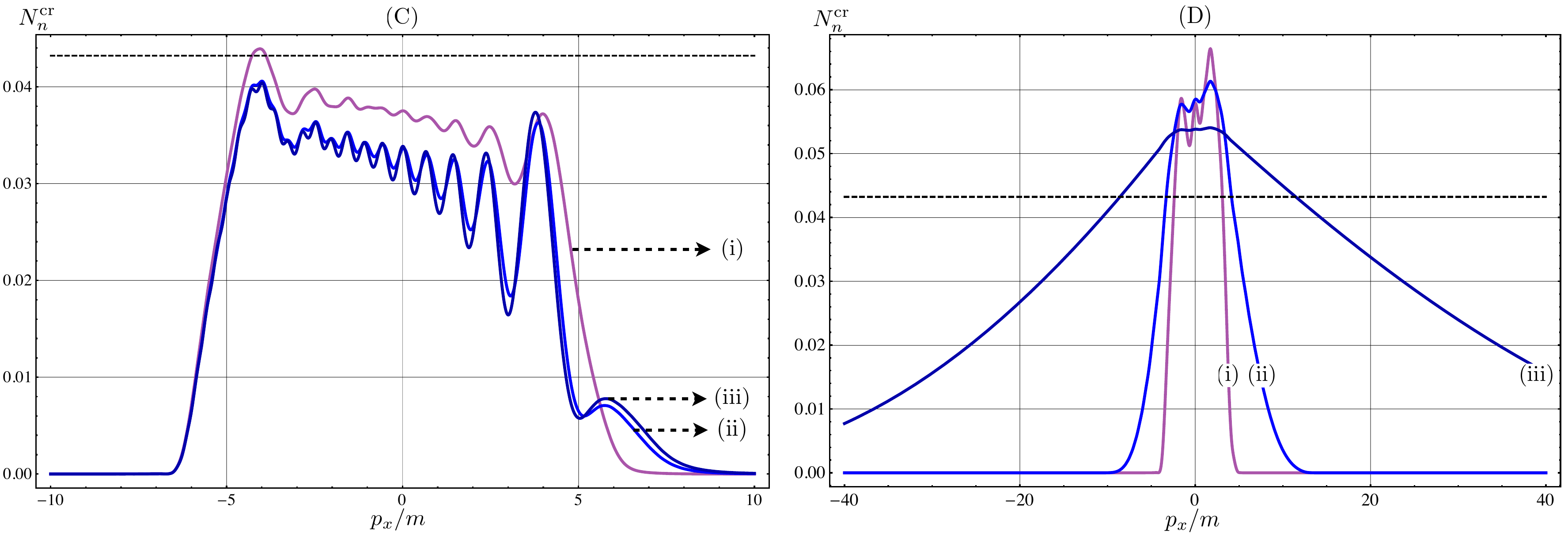}
\end{center}
\caption{(color online) Differential mean number of Bosons created from the
vacuum $N_{n}^{\mathrm{cr}}$ (solid lines) by asymmetrical composite fields,
with $k_{1}\neq k_{2}$ and amplitude $E=E_{\mathrm{c}}=m^{2}/e=1$ fixed. 
In both graphs, $mT$ is fixed, where Graph (C) shows $mT=10$ and $k_{1}/m=0.5$ while in Graph (D) shows $mT=5$.
In (C), the solid lines labeled with $(\mathrm{i})$, $(\mathrm{ii})$ and $(\mathrm{iii})$ refers to $k_2/m=1$, $k_2/m=5$ and $k_2/m=10$, respectively. In (D), $(\mathrm{i})$ denotes $k_1/m=0.5,\,k_2/m=0.3$, $(\mathrm{ii})$ denotes $k_1/m=0.1,\,k_2/m=0.07$ and $(\mathrm{iii})$ denotes $k_1/m=0.01,\,k_2/m=0.008$. The horizontal dashed line corresponds to the uniform distribution $e^{-\protect\pi \protect
\lambda }$ which, in this system of units and $\mathbf{p}_{\perp }=0$ is $e^{-\protect\pi }$.}
\label{Fig2b}
\end{figure}

Aside these properties, it is useful to visualize how the differential mean
numbers $N_{n}^{\mathrm{cr}}$ are distributed over the quantum numbers (for
instance $p_{x}$) to outline some preliminary remarks concerning pair
creation, especially in cases where asymptotic approximations of the WPCFs.
and CHFs. involved in the $g$-coefficients are not applicable\footnote{%
For example, when the argument of WPCFs. $z_{j}$ or of CHFs. $\eta _{j}$ are
finite quantities. Also when the parameters $a_{j}$, $c_{j}$ are also finite.%
}. To this end, we present below some plots of the mean number of particles
created from the vacuum $N_{n}^{\mathrm{cr}}$ (\ref{NP}) as a function of $%
p_{x}$ for different values of $k_{1}$, $k_{2}$ and $T$ (Figs. \ref{Fig1a}, %
\ref{Fig1b} for Fermions and \ref{Fig2a}, \ref{Fig2b} for Bosons) for a
fixed amplitude $E$ of the composite field. For the sake of simplicity, we
set $\mathbf{p}_{\perp }=0$ and select a convenient system of units, in
which besides $\hslash =c=1$ the electron mass is also set equal to the
unity, $m=1$.
In this system, the Compton wavelength corresponds to one unit
of length $\lambdabar _{e}=\hslash /mc=1 \approx
3.8614\times 10^{-14}\,\mathrm{m}$, the Compton time corresponds to one unit of time $%
\lambdabar _{e}/c=1 \approx 1.3\times 10^{-21}\,%
\mathrm{s}$ and the electron rest energy corresponds to one unit of energy $%
mc^{2}=1 \approx 0.511\ $\textrm{MeV}. In all plots
below, the longitudinal momentum $p_{x}$, time duration $T$and phases $k_{j}$
are relative to electron's mass $m$, corresponding to dimensionless
quantities, i. e., $p_{x}/m$, $mT$ and $k_{j}/m$, respectively.

The results displayed in all pictures above, reveal wider distributions
corresponding to composite electric fields with larger $T$ (red/dark blue
lines for Fermions/Bosons in graphs (a), Figs. \ref{Fig1a}, \ref{Fig2a}) or
smaller $k_{j}$ (red/dark blue lines for Fermions/Bosons in graphs (b),
Figs. \ref{Fig1a}, \ref{Fig2a}) and thinner distributions corresponding to
opposite configurations, associated with smaller $T$(orange/purple lines for
Fermions/Bosons in graphs (a), Figs. \ref{Fig1a}, \ref{Fig2a}) or larger $%
k_{j}$ (orange/purple lines for Fermions/Bosons in graphs (b), \ref{Fig1a}, %
\ref{Fig2a}). Once the time duration is designated by $T$ and $k_{j}^{-1}$ ($%
k_{j}^{-1}$ represent a scale of time duration for increasing and decreasing
phases), these results are consistent with the fact that the larger the
duration of an electric field, the longer it has to accelerate pairs.
Therefore larger values to $p_{x}/m$ are expected to occur in cases
corresponding to electric fields with larger time duration. Moreover, it
should be noted that the distributions above tend to the uniform
distribution $N_{n}^{\mathrm{cr}}=e^{-\pi \lambda }$ (horizontal dashed
lines) for $T$ and $k_{j}^{-1}$ sufficiently large. This is not unexpected
since the composite field tends to a constant field as soon as $T$ and $%
k_{j}^{-1}$ increase, becoming infinitely constant in the limit $%
T\rightarrow \infty $ and $k_{j}^{-1}\rightarrow \infty $. At last, but not
least, observing Figs. \ref{Fig1b}, \ref{Fig2b} we find that asymmetrical
configurations $\left( k_{1}\neq k_{2}\right) $ yields to asymmetrical
distributions. This is associated with the fact that different phases $%
k_{1},k_{2}$ implies in different times to accelerate pairs during the
switching-on and -off processes in general. An interpretation of these
results follows from the semiclassical analysis: Electrons created from the
vacuum have quantum numbers $p_{x}$ within the range $-eE\left(
T/2+k_{1}^{-1}\right) \leq p_{x}\leq eE\left( T/2+k_{2}^{-1}\right) $,
corresponding to longitudinal kinetic momenta $\Pi _{x}\left( t\right)
=p_{x}+eA_{x}\left( t\right) $ which, at $t\rightarrow +\infty $, varies
according to $-eE\left( T+k_{1}^{-1}+k_{2}^{-1}\right) \leq \Pi _{x}\left(
+\infty \right) \leq 0$. Assuming that pairs are materialized from the
vacuum with zero longitudinal momentum $\Pi _{x}\left( t\right) =0$, it
follows from the classical equations of motion that the kinetic longitudinal
momentum at $t\rightarrow +\infty $ has the form $\Pi _{x}\left( +\infty
\right) =-e\int_{t}^{+\infty }dt^{\prime }E\left( t^{\prime }\right) $,
where $t$ is the time of creation. Thus, if an electron is created at $%
t\rightarrow -\infty $, its longitudinal kinetic momentum at $t\rightarrow
+\infty $ is maximal (in absolute value) $\Pi _{x}\left( +\infty \right)
=-eE\left( T+k_{1}^{-1}+k_{2}^{-1}\right) $. At the same time, its
longitudinal kinetic momentum is expressed in terms of $p_{x}$ as $\Pi
\left( +\infty \right) =p_{x}-eE\left( T/2+k_{2}^{-1}\right) $, which means
that such a electron is found to have the a minimal value to $p_{x}$, namely
$p_{x}\rightarrow p_{x}^{\min }=-eE\left( T/2+k_{1}^{-1}\right) $. On the
other hand, if the electron is created at $t\rightarrow +\infty $, then its
longitudinal kinetic momentum tends to zero, $\Pi _{x}\left( +\infty \right)
\rightarrow 0$, which means that the corresponding quantum number $p_{x}$
tends to its maximum, $p_{x}\rightarrow p_{x}^{\max }=eE\left(
T/2+k_{2}^{-1}\right) $. According to this interpretation, asymmetric
configurations result in asymmetric distributions. This explains asymmetric
distributions in graphs (C) and (D), Figs. \ref{Fig1b}, \ref{Fig2b}, for
instance.

\section{Differential and total quantities in some special configurations
\label{Sec4}}

Irrespective of the $t$-electric potential step under consideration, it is
known that the most favorable conditions for pair creation from the vacuum
are associated with strong fields acting over a sufficiently large period of
time, in which differential and total quantities are significant. For the
composite electric field (\ref{s2.0}), the time duration is encoded in two
sets, namely, $\left( k_{1}^{-1},k_{2}^{-1}\right) $ and $\left(
t_{1},t_{2}\right) $. The former represent scales of time duration for the
increasing and decreasing phases of the electric field, defined at intervals
$\mathrm{I}$ and $\mathrm{III}$, while the latter corresponds to the time
duration in which the field is constant, defined at interval $\mathrm{II}$.

If the period $T$\ is a relatively short
(see, e.g., the cases with $mT=5$ and $k/m<0.5$ on the right side of Figs. \ref{Fig1a} - \ref{Fig2b}),
the effects of pair creation tend
to ones obtained for the peak field \cite{AdoGavGit17,AdoGavGit16}. The
latter field correspond to a limit of the composite field when the
intermediate interval $T$ is absent. From the results above, we observe that
the existence of a finite interval $T$, between \textquotedblleft
slow\textquotedblright\ switching-on and -off processes, has no significant
influence on the distribution of the differential mean numbers $N_{n}^{%
\mathrm{cr}}$ over the quantum numbers (see appropriate asymptotic formulae
in Ref. \cite{AdoGavGit17}). The influence of the $T$-constant interval
appears only in the next-to-leading order.

A composite electric field of large duration corresponds to small values for
the switching-on phase $k_{1}$, switching-off phase $k_{2}$ and large $%
T=t_{2}-t_{1}$,\footnote{%
Without loss of generality, we select from now on a symmetrical interval
\textrm{II}, in which $t_{1}=-T/2=-t_{2}$.} satisfying the following
condition%
\begin{equation}
\min \left( \sqrt{eE}T,eEk_{1}^{-2},eEk_{2}^{-2}\right) \gg \max \left( 1,%
\frac{m^{2}}{eE}\right) \,.  \label{s3.1}
\end{equation}%
The condition (\ref{s3.1}) defines a configuration in which the field takes
a sufficiently large time to reach the constant regime (slow switching-on
process, $k_{1}^{-1}$ large), remains constant over a sufficiently large
interval $T$ and finally takes another sufficiently large time to switch-off
completely (slow switching-on process, $k_{2}^{-1}$ large).

The most important objects in vacuum instability by external fields are the
total number of particles created from the vacuum $N^{\mathrm{cr}}$ and the
vacuum-to-vacuum transition probability $P_{v}$, both given by Eq. (\ref{NP}%
). The first quantity corresponds to the summation of the differential mean
numbers $N_{n}^{\mathrm{cr}}$ over the momenta $\mathbf{p}$, and spin
degrees-of-freedom%
\begin{equation}
N^{\mathrm{cr}}=V_{\left( d-1\right) }n^{\mathrm{cr}}\,,\ \ n^{\mathrm{cr}}=%
\frac{J_{\left( d\right) }}{\left( 2\pi \right) ^{d-1}}\int d\mathbf{p}%
N_{n}^{\mathrm{cr}}\,,  \label{st1}
\end{equation}%
which, in fact, is reduced to the calculation of the density of pairs
created from the vacuum $n^{\mathrm{cr}}$. Here the summation over $\mathbf{p%
}$ was transformed into an integral and $J_{\left( d\right) }=2^{\left[ d/2%
\right] -1}$ denotes the total number spin projections in a $d$-dimensional
space. These are factored out since the numbers $N_{n}^{\mathrm{cr}}$ are
independent of spin polarization. The dominant contribution of the densities
$n^{\mathrm{cr}}$ in the slowly varying regime are proportional to the total
increment of the longitudinal kinetic momentum, $\Delta U=\left\vert \Pi
_{2}-\Pi _{1}\right\vert =e\left\vert A_{x}\left( +\infty \right)
-A_{x}\left( -\infty \right) \right\vert $, which is the largest parameter
in the problem \cite{GavGit17}. Hence it is meaningful to approximate the
total density $n^{\mathrm{cr}}$ by its dominant contribution $\tilde{n}^{%
\mathrm{cr}}$, corresponding to an integral over an specific domain $\Omega$%
\begin{equation}
\Omega: n^{\mathrm{cr}}\approx \tilde{n}^{\mathrm{cr}}=\frac{J_{\left(
d\right) }}{\left( 2\pi \right) ^{d-1}}\int_{\mathbf{p}\in \Omega }d\mathbf{p%
}N_{n}^{\mathrm{cr}}\,,  \label{st1b}
\end{equation}%
whose result is proportional to $\Delta U$. As it is general for $t$%
-electric potential steps, such domain $\Omega $ is defined by a specific
range of values to the longitudinal momentum $p_{x}$ and restricted values
to the perpendicular momentum $\mathbf{p}_{\perp }$ which, under the
condition (\ref{s3.1}), is%
\begin{equation}
\Omega :\left\{ \frac{|p_{x}|}{\sqrt{eE}}\leq\sqrt{eE}\frac{T}{2}+\frac{3}{2}\sqrt{\frac{h_1}{2}}\,,\,\, \sqrt{\lambda}<K_{\perp}\,,\,\,K_{\perp}^{2}\gg\max \left(1,\frac{m^2}{eE}\right)\right\}\,. \label{st1c}
\end{equation}

In this case using asymptotic formulas given by Ref. \cite{AdoGavGit17} one
can see that the differential mean numbers are practically uniform over a
wide range of values to the kinetic momenta of the domain $\Omega $ while
decreases exponentially beyond these ranges. In leading-order approximation,
the mean numbers are%
\begin{equation}
N_{n}^{\mathrm{cr}}\sim \left\{
\begin{array}{ll}
\exp \left( -2\pi \Xi _{1}^{-}\right) \,, & \mathrm{for}\ \ p_{x}/\sqrt{eE}<-%
\sqrt{eE}T/2\,, \\
e^{-\pi \lambda }\,, & \mathrm{for}\ \ \left\vert p_{x}\right\vert /\sqrt{eE}%
\leq \sqrt{eE}T/2\,, \\
\exp \left( -2\pi \Xi _{2}^{+}\right) \,, & \mathrm{for}\ \ p_{x}/\sqrt{eE}>+%
\sqrt{eE}T/2\,.%
\end{array}%
\right.   \label{fas17}
\end{equation}%
 It is clear that the asymptotic forms
(\ref{fas17}) specified in each range above, coincides with asymptotic forms
of the $T$-constant and exponential electric fields; see, e. g., Ref. \cite%
{AdoGavGit17}. Thus, we see that in each domain of $\Omega $ with a particular
type of field, principal terms in the distribution $N_{n}^{\mathrm{cr}}$ do
not depend on the type of fields in neighboring regions, only terms of
following orders acquire such a dependence. It follows that\emph{\ }the
dominant contribution for the density of pairs created by the composite
field is expressed as a sum of the dominant contribution for the $T$%
-constant and exponential electric fields,%
\begin{equation}
\tilde{n}^{\mathrm{cr}}\approx \sum_{j}\tilde{n}_{j}^{\mathrm{cr}},\ \
\tilde{n}_{j}^{\mathrm{cr}}=\frac{J_{\left( d\right) }}{\left( 2\pi \right)
^{d-1}}\int_{t\in D_{j}}dt\left[ eE_{j}\left( t\right) \right] ^{d/2}\exp %
\left[ -\pi \frac{m^{2}}{eE_{j}\left( t\right) }\right] \,,  \label{st13c}
\end{equation}%
where the index $j=1,2,3$ denotes each interval of the composite field, $%
D_{1,2,3}=\mathrm{I,II,III}$. It is known \cite{AdoGavGit17} that%
\begin{eqnarray}
\tilde{n}^{\mathrm{cr}}_{1,3} &=&\frac{J_{\left( d\right) }}{\left( 2\pi
\right) ^{d-1}}\frac{\left( eE\right) ^{d/2}}{k_{1,2}}e^{-\pi
m^{2}/eE}G\left( \frac{d}{2},\frac{\pi m^{2}}{eE}\right) \,.  \notag \\
\tilde{n}^{\mathrm{cr}}_{2} &=&\frac{J_{\left( d\right) }\left( eE\right)
^{d/2}T}{\left( 2\pi \right) ^{d-1}}\exp \left[ -\frac{\pi m^{2}}{eE}\right]
\,,  \label{st13d}
\end{eqnarray}%
where $G\left( \alpha ,z\right) $ is expressed in terms of the incomplete
gamma function $\Gamma \left( \alpha ,z\right) $ \cite{DLMF} as%
\begin{equation}
G\left( \alpha ,z\right) =\int_{1}^{\infty }\frac{ds}{s^{\alpha +1}}%
e^{-z\left( s-1\right) }=e^{z}z^{\alpha }\Gamma \left( -\alpha ,z\right) \,.
\label{st7}
\end{equation}%
Calculating the vacuum-to-vacuum probability of the composite field, we
obtain that it is product of the partial $P_{v}^{j}$ for the $T$-constant
and exponential electric fields, respectively, $\ln P_{v}=\sum_{j}\ln
P_{v}^{j}$; see the Ref. \cite{AdoGavGit17}. It is important to point out
that the result above may be reproduced from the universal form for the
total density of pairs created by $t$-electric potential steps in the slowly
varying regime \cite{GavGit17}. Such a form does not demand knowledge on the
exact solutions of the Dirac/Klein-Gordon equations. This is a consequence
of the fact that in the approximation by leading terms, the distribution $%
N_{n}^{\mathrm{cr}}$ in each region of $\Omega $ is formed independently of
neighboring regions.

While the results for slowly varying fields are completely predictable,
configurations in which fields act over a relatively short time to reach the
constant regime (fast switching-on process, $k_{1}^{-1}$ small), remain
constant over a sufficiently large interval $T$ and takes a short interval
to switch-off completely (fast switching-off process, $k_{2}^{-1}$ small)
have to be studied in more details. These configurations simulate finite
switching-on and -off processes, whose considerations are discussed below.

To study such configurations, one has to compare parameters involving
momenta with ones involving time scales, such as $\sqrt{eE}T$, $eEk_{1}^{-2}$
and $eEk_{2}^{-2}$. Regarding the dependence on the perpendicular momenta $%
\mathbf{p}_{\perp }$ for instance, it is well known that a $t$-electric
potential step of large time duration does not create a significant number
of pairs with large $\mathbf{p}_{\perp } $. This is meaningful as long as
charged pairs are accelerated along the direction of the electric field,
having thereby a wider range of values of $p_{x}$ instead $\mathbf{p}_{\perp
}$. By virtue of that, one may simplify the calculation of differential
quantities and consider restricted values to $\mathbf{p}_{\perp }$, ranging
from zero till a finite number so that the inequality%
\begin{equation}
\sqrt{\lambda }<K_{\perp }\,,\ \ K_{\perp }^{2}\gg \max \left( 1,\frac{m^{2}%
}{eE}\right) \,,  \label{s3.2}
\end{equation}%
is fulfilled. Here $K_{\perp }$ is a moderately large number that sets an
upper bound to the perpendicular momenta of pairs created. Thus, taking into
account the inequality above, we assume that
\begin{equation}
\sqrt{eE}T\gg K_{\perp }^{2}\,,\ \ \max \left(
eEk_{1}^{-2},eEk_{2}^{-2}\right) \leq \max \left( 1,\frac{m^{2}}{eE}\right)
\,,  \label{s3.3}
\end{equation}%
As a consequence, the field satisfies the following inequalities%
\begin{equation}
\sqrt{eE}T/2\gg \max \left( \sqrt{eE}k_{1}^{-1},\sqrt{eE}k_{2}^{-1}\right)
\leftrightarrow \max \left( k_{1}T/2,k_{2}T/2\right) \gg 1\,.  \label{s3.3.1}
\end{equation}

To study differential quantities in this case we select a definite sign for $%
p_{x}$ which, for convenience, the negative is chosen $-\infty <p_{x}\leq 0$%
. Next we use the properties of symmetry discussed in Eqs. (\ref{sym1}) and (%
\ref{sym2}) to generalize results for $p_{x}$ positive. Here $\xi _{1}$
varies from large negative to large positive values while $\xi _{2}$ is
always large and positive; $\Pi _{1}/\sqrt{eE}$ changes from large positive
to large negative values while $\Pi _{2}/\sqrt{eE}$ that is always large and
negative. However once $h_{1}$,$h_{2}$ are finite, we find that the
asymptotic behavior of $N_{n}^{\mathrm{cr}}$ is classified according to
three main ranges%
\begin{eqnarray}
&&\left( \mathrm{a}\right) \ \ -\sqrt{eE}\frac{T}{2}\leq \xi _{1}\leq -%
\tilde{K}_{1}\leftrightarrow \sqrt{eE}\frac{T}{2}+\sqrt{\frac{h_{1}}{2}}\geq
\frac{\Pi _{1}}{\sqrt{eE}}\geq \tilde{K}_{1}+\sqrt{\frac{h_{1}}{2}}\,,
\notag \\
&&\left( \mathrm{b}\right) \ \ -\tilde{K}_{1}<\xi _{1}<\tilde{K}%
_{1}\leftrightarrow \tilde{K}_{1}+\sqrt{\frac{h_{1}}{2}}>\frac{\Pi _{1}}{%
\sqrt{eE}}>-\tilde{K}_{1}+\sqrt{\frac{h_{1}}{2}}\,,  \notag \\
&&\left( \mathrm{c}\right) \ \ \xi _{1}\geq \tilde{K}_{1}\leftrightarrow
\frac{\Pi _{1}}{\sqrt{eE}}\leq -\tilde{K}_{1}+\sqrt{\frac{h_{1}}{2}}\,,
\label{fas19}
\end{eqnarray}%
where $\tilde{K}_{1}$ is a sufficiently large number satisfying $\sqrt{eE}T>%
\tilde{K}_{1}\gg K_{\perp }^{2}$. Moreover, as long as $\xi _{2}$ is large
and positive, $c_{2}$ is also large so that one case use the asymptotic
approximation (9.246.1) in Ref. \cite{Gradshteyn} for the WPCfs. $u_{\pm
}\left( z_{2}\right) $ and Eq. (13.8.2) in Ref. \cite{DLMF} for the CHF $%
y_{1}^{2}\left( \eta _{2}\right) $, throughout all ranges above.

In the range $\left( \mathrm{a}\right) $, $\xi _{1}$ is large and negative
and $c_{1}$ is large as well. Then using the asymptotic expansions
(9.246.2), (9.246.3) in Ref. \cite{Gradshteyn} for the WPCfs. $u_{\pm
}\left( z_{1}\right) $ and Eq. (13.8.2) in \cite{DLMF} for the CHF $%
y_{2}^{1}\left( \eta _{1}\right) $, one finds that the mean number of
particles created, in the leading order approximation, admit the following
form%
\begin{eqnarray}
N_{n}^{\mathrm{cr}} &\sim &\frac{\exp \left[ -\pi \left( \lambda +\Xi
_{1}^{-}-\Xi _{2}^{+}\right) \right] }{\sinh \left( 2\pi \omega
_{2}/k_{2}\right) \sinh \left( 2\pi \omega _{1}/k_{1}\right) }  \notag \\
&\times &\left\{
\begin{array}{ll}
\sinh \left( \pi \Xi _{2}^{-}\right) \sinh \left( \pi \Xi _{1}^{+}\right) \,,
& \mathrm{Fermi} \\
\cosh \left( \pi \Xi _{2}^{-}\right) \cosh \left( \pi \Xi _{1}^{+}\right) \,,
& \mathrm{Bose}%
\end{array}%
\right. \,,  \label{fas20}
\end{eqnarray}%
as $T\rightarrow \infty $. The combination of hyperbolic functions above
tends to the unity since, in this range, the frequencies $\omega _{1}$, $%
\omega _{2}$ and the parameters $\Xi _{1}^{+}$, $\Xi _{2}^{-}$ are large
quantities, namely $\omega _{1}\simeq \sqrt{eE}\left\vert \xi
_{1}\right\vert $, $\omega _{2}\simeq \sqrt{eE}\xi _{2}$, $\Xi _{1}^{+}\sim
\sqrt{2h_{1}}\left\vert \xi _{1}\right\vert $, $\Xi _{2}^{-}\simeq \sqrt{%
2h_{2}}\xi _{2}$. In virtue of that, the dominant contribution of Eq. (\ref%
{fas20}) has the form%
\begin{equation}
N_{n}^{\mathrm{cr}}\sim \exp \left[ -\pi \left( \lambda +2\Xi
_{1}^{-}\right) \right] \,,  \label{fas21}
\end{equation}%
as $T\rightarrow \infty $, being valid both for Fermions and Bosons. In this
last result, the parameter $\Xi _{1}^{-}$ is a small quantity, $\Xi
_{1}^{-}\sim \sqrt{h_{1}/2}\left( \lambda /2\left\vert \xi _{1}\right\vert
\right) $, so that its contribution to $N_{n}^{\mathrm{cr}}$ are negligible
in comparison to $\lambda $. As a result, the differential mean numbers are
practically uniform over the range $\left( \mathrm{a}\right) $, $N_{n}^{%
\mathrm{cr}}\sim e^{-\pi \lambda }$.

In the range $\left( \mathrm{c}\right) $, $\xi _{1}$ is large and positive
and $c_{1}$ is also large. Hence one may use the asymptotic expansions
(9.246.1) in Ref. \cite{Gradshteyn} for the WPCfs. $u_{\pm }\left(
z_{1}\right) $ and Kummer transformations for the CHF $y_{2}^{1}\left( \eta
_{1}\right) $ to prove that the mean number of particles created is
significantly small%
\begin{equation}
N_{n}^{\mathrm{cr}}\sim \mathcal{F}_{1}\left[ O\left( \xi _{1}^{-6}\right)
+O\left( \xi _{2}^{-6}\right) +O\left( \xi _{1}^{-3}\xi _{2}^{-3}\right) %
\right] \,,  \label{fas22}
\end{equation}%
as $T\rightarrow \infty $, in which $\mathcal{F}_{1}$ is a combination of
hyperbolic functions similar to Eq. (\ref{fas20}),%
\begin{eqnarray}
\mathcal{F}_{1} &=&\frac{\exp \left[ \pi \left( \Xi _{2}^{+}+\Xi
_{1}^{+}\right) \right] }{\sinh \left( 2\pi \omega _{2}/k_{2}\right) \sinh
\left( 2\pi \omega _{1}/k_{1}\right) }  \notag \\
&\times &\left\{
\begin{array}{ll}
\sinh \left( \pi \Xi _{2}^{-}\right) \sinh \left( \pi \Xi _{1}^{-}\right) \,,
& \mathrm{Fermi\,,} \\
\cosh \left( \pi \Xi _{2}^{-}\right) \cosh \left( \pi \Xi _{1}^{-}\right) \,,
& \mathrm{Bose\,.}%
\end{array}%
\right.  \label{fas23}
\end{eqnarray}%
In this range, the frequencies $\omega _{j}$ and the parameters $\Xi
_{j}^{-} $ are large quantities $\omega _{j}\simeq \sqrt{eE}\xi _{j}$, $\Xi
_{j}^{-}\sim \sqrt{2h_{j}}\xi _{j}$ so that, as in the range $\left( \mathrm{%
a}\right) $, $\mathcal{F}_{1}$ can be approximated to $\mathcal{F}_{1}\sim 1$%
. Therefore the differential mean numbers are significantly small in this
range.

In the range $\left( \mathrm{b}\right) $, $\xi _{1}$ varies from large
negative to large positive values while $c_{1}$ varies from large to finite
values. By this reason, it is not possible to use any asymptotic
approximations for the special functions $u_{\pm }\left( z_{1}\right) $ and $%
y_{2}^{1}\left( \eta _{1}\right) $, although one can still consider the same
approximations (9.246.1) in Ref. \cite{Gradshteyn} and (13.8.2) in Ref. \cite%
{DLMF} for the WPCfs. $u_{\pm }\left( z_{2}\right) $ and CHF $%
y_{1}^{2}\left( \eta _{2}\right) $, respectively. The resulting expression
shall depend explicitly on the exact forms of $u_{\pm }\left( z_{1}\right) $
and $y_{2}^{1}\left( \eta _{1}\right) $.

The most significant contribution for the differential mean numbers for $%
p_{x}$ positive, $0\leq p_{x}<+\infty $, can be obtained by a similar
analysis, taking into account the properties of symmetry (\ref{sym1}) and (%
\ref{sym2}). We finally find the domain of dominant contribution to the mean
number of particles created. In this domain in the leading order
approximation, it has a form

\begin{equation}
N_{n}^{\mathrm{cr}}\sim e^{-\pi \lambda }\times \left\{
\begin{array}{ll}
\exp \left( -2\pi \Xi _{1}^{-}\right) \,, & \mathrm{for}\ \ -\sqrt{eE}T/2+%
\tilde{K}_{1}<p_{x}/\sqrt{eE}\leq 0\,, \\
\exp \left( -2\pi \Xi _{2}^{+}\right) \,, & \mathrm{for}\ \ 0<p_{x}/\sqrt{eE}%
\leq \sqrt{eE}T/2-\tilde{K}_{2}\,,%
\end{array}%
\right.  \label{fas27}
\end{equation}%
as $T\rightarrow \infty $, valid for Fermions and Bosons. This approximation
is almost uniform over this wide range of values to the longitudinal
momentum since the parameters $\Xi _{1}^{-}$ and $\Xi _{2}^{+}$ are
negligible in comparison to $\lambda $. In virtue of that, the switching-on
and -off effects on the differential mean numbers, in the present
configuration, manifest themselves as next-to-leading corrections to the
uniform distribution $e^{-\pi \lambda }$. This means that the influence of
the switching-on and -off processes on differential quantities are
negligible for $T$ sufficiently large. From these results, the present
configuration can be referred as a \textquotedblleft fast\textquotedblright\
switching-on and -off configuration, in virtue of  Eq. (%
\ref{s3.3.1}) and from the fact that the mean number of particles created
are mainly characterized by the uniform distribution $e^{-\pi \lambda }$. In
this case, the leading contribution to the number density $\tilde{n}^{%
\mathrm{cr}}$, given by Eq.~(\ref{st1b}),\ is proportional to the total
increment of the longitudinal kinetic momentum,\ $\Delta U=eET\ $, and then
the time duration $T$. We see that both the $T$-constant field
itself and the composite field under condition (\ref{s3.3.1}) can be
considered as regularizations of a constant field. The present
discussion encompasses the $T$\ -constant limit, characterized by the
absence of exponential parts and defined by the limit $k\rightarrow \infty $.

\begin{figure}[th]
\begin{center}
\includegraphics[scale=0.48]{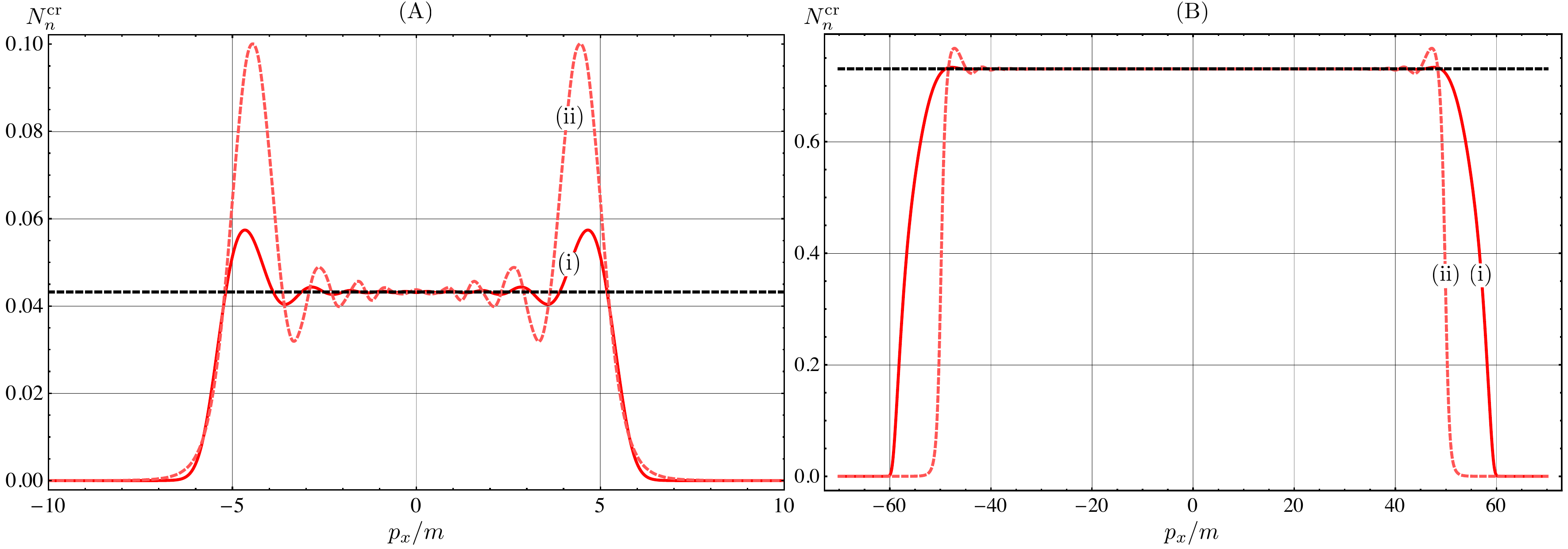}
\end{center}
\caption{(color online) Differential mean number of electron/positron pairs
created from the vacuum by a symmetric composite field (solid red lines, labeled with (i))
with $k_{1}/m=k_{2}/m=1$ and by a $T$-constant field (dashed light red
lines, labeled with (ii)). In panel (A), $E=E_{\mathrm{c}}$ while in  panel (B), $E=10E_{\mathrm{c}}$. In both cases, $mT=10$ and $\mathbf{p}_{\perp }=0$%
. The horizontal dashed black line denotes the uniform distributions, being $%
e^{-\protect\pi } $ in (A) and $e^{-\protect\pi /10}$ in (B).}
\label{Fermi}
\end{figure}

\begin{figure}[th]
\begin{center}
\includegraphics[scale=0.48]{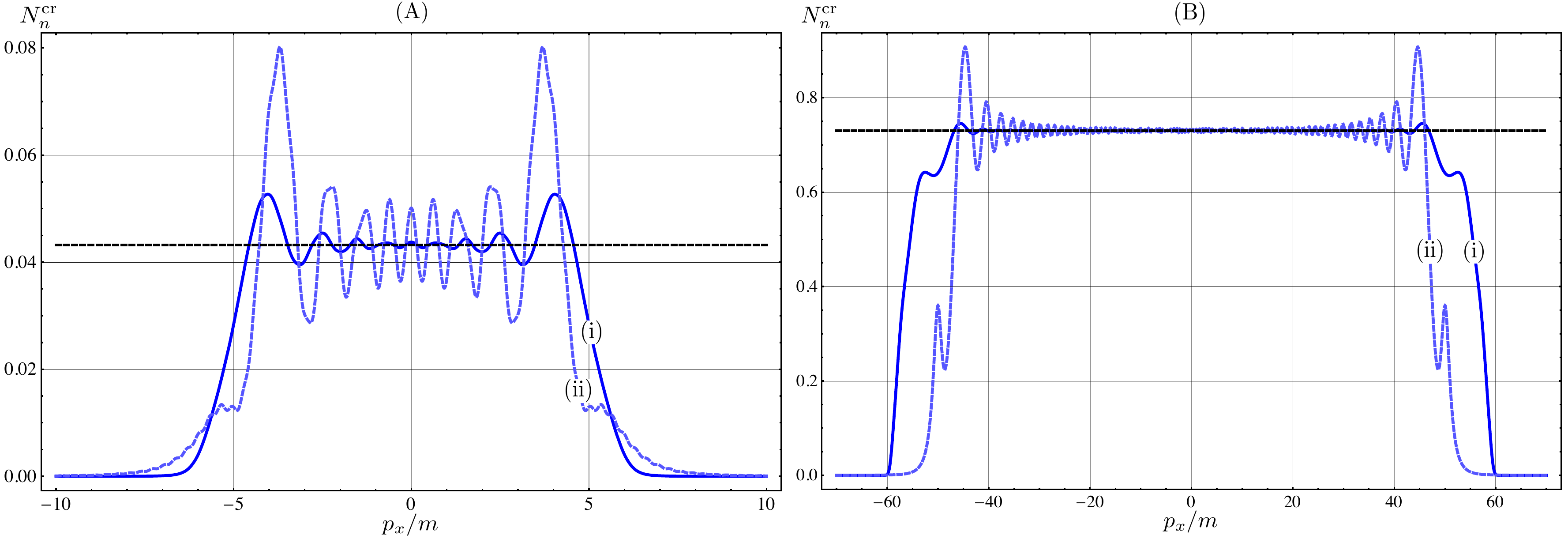}
\end{center}
\caption{(color online) Differential mean number of scalar particles created
from the vacuum by a symmetric composite field (solid blue lines, labeled with (i)) with $%
k_{1}/m=k_{2}/m=1$ and by a $T$-constant field (dashed light blue lines, labeled with (ii)). In
panel (A), $E=E_{\mathrm{c}}$ while in panel (B), $%
E=10E_{\mathrm{c}}$. In both cases, $mT=10$ and $\mathbf{p}_{\perp }=0$. The
horizontal dashed black line denotes the uniform distributions, $e^{-\protect%
\pi }$ in (A) and $e^{-\protect\pi /10}$ in (B).}
\label{Bose}
\end{figure}

We know that a possibility of describing particle creation by the $T$%
-constant field in the slowly varying approximation depends of the value of
dimensionless parameter $\sqrt{eE}T>1.$ According to condition (\ref{s3.3})
a magnitude of the lower boundary $\vartheta =\min \sqrt{eE}T$\ is
proportional to $m^{2}/eE$ if $m^{2}/eE>1$. Accordingly, a contribution of
switching-on and -off processes to the particle creation effect becomes more
pronounced for not too strong fields. It is useful to compare switching-on
and -off effects for the $T$-constant field and for the composite field in
the case when the parameter $\sqrt{eE}T$ approaches the above mentioned
threshold values.
From the plots on the left side of Figs. \ref{Fig1a} - \ref{Fig2b}) one can see that $\sqrt{eE}T=10$ is near threshold value.
 To this end we compute exact plots of the mean
differential number of Fermions (\ref{r4}) and Bosons (\ref{r7}) created as
a function of $p_x/m$ for two typical cases of critical field and very
strong field, respectively. In the case of the $T$-constant field, we
calculate the $p_{x}/m$ dependence using exact Eqs. (4.9) and (4.11) given
in Ref. \cite{AdoGavGit17}. Results of these computations are presented on
Figs. \ref{Fermi} and \ref{Bose}. We see that differential mean numbers of
pairs created by the composite electric field (solid lines) and the $T$%
-constant field (dashed lines) oscillate around the uniform distribution $%
e^{-\pi \lambda }$. It can be seen that for fields with a critical
magnitude, $E=E_{\mathrm{c}}$\ and $\sqrt{eE}T=10$ (plots (a)), the
oscillations around the uniform distribution are greater than for fields
with overcritical magnitude, $E=10E_{\mathrm{c}}$\ and $\sqrt{eE}T=10\sqrt{10%
}$\ (plots (b)), both for the composite field and the $T$-constant field.

We see that distributions $N_{n}^{\mathrm{cr}}$ for the $T$-constant field
always oscillate greater around the uniform distribution than for the
composite field. And in the case of bosons, these deviations from the
uniform distribution are more significant. On the other hand, the plot of $%
N_{n}^{\mathrm{cr}}$\ for the $T$-constant field is more \textquotedblleft
rectangular\textquotedblright\ than for the composite field (for
overcritical magnitudes).\emph{\ }Such wide distributions arose because of
contributions of exponential tails,\emph{\ }$\left\vert p_{x}\right\vert
/m<\left( \frac{eE}{m^{2}}\right) \left( \frac{mT}{2}+\frac{m}{k}\right) $.
 Note also that for $\left\vert p_{x}\right\vert /m>\left( \frac{eE}{%
m^{2}}\right) \left( \frac{mT}{2}+\frac{m}{k}\right) $, mean numbers for the
composite field for both magnitudes are negligible, whereas for the $T$%
-constant field this is not always true: in fact, for critical magnitudes,
the mean numbers for $\left\vert p_{x}\right\vert /m$ slightly larger than $%
\left( \frac{eE}{m^{2}}\right) \left( \frac{mT}{2}\right) $ are not
negligible, although they are small.\emph{\ }The characteristic behavior$\ $%
in the case of the slowly varying regime, when$\ \tilde{n}^{\mathrm{cr}}\sim
T,\ $is quite noticeable in the case of fermions already for the value of
the dimensionless parameter$\ \sqrt{eE}T=10\ $and is pronounced for large
values of this parameter.\emph{\ }It can be concluded that for fermions the
quantity$\ \sqrt{eE}T\ =\ 10\ $is close to the threshold value. However, for
bosons at$\ \sqrt{eE}T\ =\ 10,\ $the approximation of the slowly varying
regime does not work yet.\emph{\ }To be applicable this approximation
requires larger values of the parameter $\sqrt{eE}T$. The slowly varying
regime is working for both fermions and bosons at $\sqrt{eE}T=10\sqrt{10}$.
Comparing these two cases,\ we\ see\ that\ the regularization by switching\
on\ and\ off\ exponential\ fields is less\emph{\ }disturbing\emph{\ }than by
the\ $T$-constant field,\ which entails considerable oscillations in the
distributions, and can even lead to sharp bursts $N_{n}^{\mathrm{cr}}$\ in
narrow regions of $p_{x}$. The latter circumstance, however, is not
essential for estimating of dominant contributions for the density of
created pairs due to the very strong $T$-constant field. However, the above
calculation method which is using the composite field is more realistic and
preferable for the analysis of next-to-leading terms.

\section{Concluding remarks\label{conclusions}}

We find exact formulas for differential mean numbers of fermions and bosons
created from the vacuum due to the composite electric field of special
configuration that simulate finite switching-on and -off processes within
and beyond the slowly varying regime. We show that the results for slowly
varying fields are completely predictable using recently developed version
of a locally constant field approximation. Using exact results beyond the
slowly varying regime, we find that the leading contribution to the number
density of created pairs is independent of fast switching-on and -off if the
time duration $T$\ of a slowly varying field is sufficiently large. It means
that composite fields of such configurations can be used as regularizations
of a slowly varying field, in particular, of a constant field.\emph{\ }We
have studied effects of fast switching-on and -off in a number of cases,
when the value of the total increment of the longitudinal kinetic momentum,
characterized by the dimensionless parameter $\sqrt{eE}T>1$,\ approaches the
threshold that determines the transition from a regime that is sensitive to
parameters of on-off switching to the slowly varying regime. It is shown
that for bosons this threshold value is much higher.\emph{\ }We\ see\ that\
the regularization by faster switching\ on\ and\ off\ is more disturbing%
\emph{,\ }which entails considerable oscillations in distributions, and can
even lead to sharp bursts $N_{n}^{\mathrm{cr}}$\ in narrow regions of $p_{x}$%
\emph{. }The latter circumstance, however, is not essential for estimating
of dominant contributions to the density of created pairs due to the very
strong field. However, the above calculation method which is using the
composite field is more realistic and preferable for the analysis of
next-to-leading terms. Thus, details of switching-on and -off may be
important for a more complete description of the vacuum instability in some
physical situations, for example, in physics of low dimensional systems,
such as graphene and similar nanostructures, whose transport properties may
be interpreted as pair creation effects under low energy approximations.

\section*{Acknowledgements}

The reported study was partially funded by RFBR according to the research
project No. 18-02-00149. The authors acknowledge support from Tomsk State
University Competitiveness Improvement Program. D.M.G. is also supported by
Grant No. 2016/03319-6, Funda\c{c}\~{a}o de Amparo \`{a} Pesquisa do Estado
de S\~{a}o Paulo (FAPESP), and permanently by Conselho Nacional de
Desenvolvimento Cient\'{\i}fico e Tecnol\'{o}gico (CNPq), Brazil.

\end{document}